\documentclass{webofc}

\usepackage[varg]{txfonts}   
\usepackage{hyperref}
\usepackage{url}
\hypersetup{colorlinks=true,citecolor=blue,urlcolor=blue,linkcolor=blue}

\title{Neutron Stars and Neutron Skins: Connecting Finite Nuclei to Dense Matter}

\author{C.A. Bertulani\inst{1}\fnsep\thanks{\email{carlos.bertulani@etamu.edu}}}

\institute{Department of Physics, East Texas A\&M University, Commerce, Texas 75429, USA}

\begin{document}

\abstract{
This is a brief overview of the connection between neutron skin thickness in finite nuclei and the equation of state of neutron-rich matter, with applications to neutron stars. Multiple experimental probes are discussed, including dipole polarizability, parity-violating electron scattering, heavy-ion fragmentation, quasi-free scattering, and ultraperipheral collisions. A consistent picture emerges from Bayesian analyses combining experimental data and energy density functionals, providing constraints on the symmetry energy and its slope.
}

\maketitle

\section{Introduction}

Nuclear physics spans a wide range of scales, from keV energies relevant to stellar evolution to TeV energies encountered in relativistic heavy-ion collisions. The nuclear many-body problem remains one of the most challenging problems in physics, largely due to the complexity of nucleon-nucleon interactions and the composite structure of nucleons themselves. 
A central objective of modern nuclear physics is to understand the properties of nuclear matter under extreme conditions. This includes its role in stellar evolution and nucleosynthesis, its behavior in the early universe, and its manifestation in compact astrophysical objects such as neutron stars. Experimental facilities such as FAIR and the future Electron-Ion Collider (EIC) provide unique opportunities to probe these regimes across a wide range of energies and densities.


\section{Neutron Stars and the EOS}

Neutron stars are compact remnants of core-collapse supernovae, characterized by masses of the order of $1-2\,M_\odot$ and radii of approximately $10-15$ km. Their internal structure is governed by the properties of matter at supra-nuclear densities, making them unique laboratories for testing the equation of state (EOS) of dense nuclear matter.
The equilibrium structure of a neutron star is determined by the balance between gravitational attraction and internal pressure. In the framework of general relativity, this balance is described by the Tolman-Oppenheimer-Volkoff (TOV) equations, which generalize the Newtonian hydrostatic equilibrium equation to relativistic systems. The TOV equations are given by
\begin{equation}
\frac{dP(r)}{dr} = -\frac{G}{r^2}
\frac{\left[\varepsilon(r) + P(r)/c^2\right]
\left[M(r) + 4\pi r^3 P(r)/c^2\right]}
{1 - 2GM(r)/(rc^2)},
\end{equation}
and
$
{dM(r)}/{dr} = 4\pi r^2 \varepsilon(r),
$
where $P(r)$ is the pressure, $\varepsilon(r)$ is the energy density, and $M(r)$ is the enclosed gravitational mass within a radius $r$. 
These coupled differential equations must be solved simultaneously, starting from a central density $\varepsilon_c$ with the boundary conditions $M(0)=0$ and $P(0)=P_c$. The integration proceeds outward until the pressure vanishes, $P(R)=0$, which defines the stellar radius $R$. The corresponding value $M(R)$ gives the total mass of the neutron star.

A key feature of the TOV equation is the presence of relativistic corrections absent in the Newtonian case. In particular, the pressure contributes to the gravitational field through the term $P/c^2$, and the factor $(1 - 2GM/rc^2)^{-1}$ accounts for spacetime curvature. These effects become essential at the high densities typical of neutron star interiors.
The EOS, providing the relation $P(\varepsilon)$, is the crucial input needed to close the system of equations. Different assumptions about the composition of dense matter, such as nucleonic matter, hyperons, or deconfined quarks, lead to different EOS predictions, which in turn produce distinct mass-radius relations. Observational constraints, including precise mass measurements and radius determinations from X-ray observations, impose stringent limits on the allowed EOS.
Therefore, determining the EOS of neutron-rich matter is a central problem that connects nuclear physics experiments with astrophysical observations. In particular, properties of finite nuclei, such as neutron skin thickness, provide valuable constraints on the symmetry energy and its density dependence, which directly influence neutron star structure.

\section{Symmetry Energy}

The energy per nucleon of asymmetric nuclear matter can be expanded around saturation density $\rho_0$ as
\begin{equation}
\frac{E}{A}(\rho,\delta) =
\frac{E}{A}(\rho_0) +
\frac{K_\infty}{18}\left(\frac{\rho-\rho_0}{\rho_0}\right)^2
+ S(\rho)\delta^2,
\end{equation}
where $\delta = (\rho_n-\rho_p)/\rho$ is the isospin asymmetry.
The symmetry energy $S(\rho)$ can be expressed as a density expansion,
\begin{equation}
S(\rho) = J + Lx + \frac{1}{2}K_{\text{sym}}x^2,
\end{equation}
with
$
x = ({\rho - \rho_0})/({3\rho_0}).
$
The slope parameter,
$
L = 3\rho_0 \left.({dS}/{d\rho})\right|_{\rho_0},
$
plays a particularly important role in determining neutron-rich matter properties and neutron star radii.

A quantitative connection between finite nuclei and the symmetry energy is most naturally established within the framework of nuclear energy density functionals (EDFs). The total energy of a nucleus is expressed as a functional of neutron and proton densities, $\rho_n(\bf{r})$ and $\rho_p(\bf{r})$, and their gradients,
\begin{equation}
E[\rho_n,\rho_p] = \int d^3r \, \mathcal{H}\big(\rho_n,\rho_p,\nabla\rho_n,\nabla\rho_p,\ldots\big),
\end{equation}
where $\mathcal{H}$ is the energy density.
For uniform nuclear matter, gradient terms vanish and the EDF reduces to a function of the total density $\rho=\rho_n+\rho_p$ and the isospin asymmetry $\delta=(\rho_n-\rho_p)/\rho$. Expanding the energy per particle, one obtains
\begin{equation}
\frac{E}{A}(\rho,\delta) = \frac{E}{A}(\rho,0) + S(\rho)\,\delta^2 + \mathcal{O}(\delta^4),
\end{equation}
which defines the symmetry energy $S(\rho)$ microscopically within the EDF.
In Skyrme-type EDFs, the symmetry energy can be written explicitly in terms of the coupling constants,
\begin{align}
S(\rho) &= \frac{1}{3}\frac{\hbar^2}{2m}\left(\frac{3\pi^2\rho}{2}\right)^{2/3}
+ \frac{1}{8}t_0(2x_0+1)\rho \nonumber \\
&\quad + \frac{1}{48}t_3(2x_3+1)\rho^{\alpha+1}
+ \frac{1}{24}\left[3t_1x_1 - t_2(4+5x_2)\right]\left(\frac{3\pi^2\rho}{2}\right)^{2/3}\rho,
\end{align}
where $t_i$, $x_i$, and $\alpha$ are Skyrme parameters. This expression illustrates how different EDF parameterizations lead to varying density dependences of $S(\rho)$, particularly away from saturation density.
Similarly, relativistic mean-field (RMF) models provide the symmetry energy through meson-exchange interactions. In these models,
\begin{equation}
S(\rho) = \frac{k_F^2}{6\sqrt{k_F^2 + m^{*2}}} + \frac{g_\rho^2}{8m_\rho^2}\rho,
\end{equation}
where $k_F$ is the Fermi momentum, $m^*$ is the effective nucleon mass, and $g_\rho$ and $m_\rho$ are the coupling and mass of the isovector $\rho$ meson.

A key outcome of EDF studies is the emergence of robust correlations between symmetry energy parameters and observables in finite nuclei. In particular, the neutron skin thickness $\Delta r_{np}$ exhibits a nearly linear correlation with the slope parameter $L$,
$
\Delta r_{np} \simeq a + b\,L,
$
with coefficients $a$ and $b$ depending weakly on the nuclear mass region. This correlation arises because both $\Delta r_{np}$ and $L$ are controlled by the pressure of neutron-rich matter near saturation density.
Microscopic EDF calculations, calibrated to nuclear masses, radii, and excitations, therefore provide a bridge between laboratory observables and the EOS of neutron-rich matter. Systematic studies employing large sets of EDFs, including both Skyrme and RMF models, have demonstrated that uncertainties in $S(\rho)$ and its derivatives can be constrained by combining data on neutron skins, dipole polarizabilities, and collective excitations.
These connections have been extensively discussed in the literature, see e.g. Refs.~\cite{Aumann2020,RocaMaza2011,Reinhard2010,Horowitz2014}, and form the basis for modern Bayesian analyses that extract symmetry energy parameters from diverse experimental datasets.


\section{Neutron Skin Thickness}

The neutron skin thickness is defined as the difference between the neutron and proton root-mean-square radii,
$
\Delta r_{np} = R_n - R_p.
$
It provides a direct measure of the isovector structure of nuclei and establishes a fundamental link between finite nuclear systems and the properties of neutron-rich matter.
From a physical standpoint, the formation of a neutron skin can be understood as a consequence of the competition between surface tension and the pressure generated by the symmetry energy. In neutron-rich nuclei, the excess neutrons experience a repulsive isovector interaction that tends to push them toward the nuclear surface. At the same time, the surface energy favors a uniform distribution of nucleons. The resulting balance determines the spatial separation between neutron and proton distributions.

A key insight emerging from both microscopic and macroscopic models is that the neutron skin thickness is strongly correlated with the density dependence of the symmetry energy, particularly with its slope parameter $L$. This correlation can be understood by considering the pressure of pure neutron matter at densities near saturation,
$
P_{\rm nm}(\rho_0) \simeq {\rho_0 L}/{3},
$
which follows directly from the definition of $L$. A larger value of $L$ corresponds to a stiffer symmetry energy, implying a larger pressure in neutron-rich matter. This increased pressure tends to push neutrons outward, leading to a thicker neutron skin.

Within the droplet model framework, this connection can be made more quantitative. The neutron skin thickness can be expressed as \cite{Reinhard2010,RocaMaza2011}
\begin{equation}
\Delta r_{np} \approx \sqrt{\frac{3}{5}} \left[
t - \frac{e^2 Z}{70 J}
+ \frac{5}{2R}(a_n^2 - a_p^2)
\right],
\end{equation}
where $t$ is the displacement between neutron and proton mean surfaces, $R$ is the nuclear radius, $J$ is the symmetry energy at saturation, and $a_n$ and $a_p$ are the surface diffuseness parameters for neutrons and protons, respectively. The dominant contribution arises from the term $t$, which depends on the symmetry energy and its density dependence.
In particular, one finds that $t$ is proportional to the ratio of the surface and bulk symmetry energy contributions and depends explicitly on $L$. To leading order, this leads to an approximately linear relationship between the neutron skin thickness and the slope parameter,
$
\Delta r_{np} \simeq a + b\,L.
$
The coefficients $a$ and $b$ encapsulate nuclear structure effects and depend weakly on the mass number and the details of the nuclear model. Physically, $a$ represents the baseline contribution to the neutron skin arising from finite-size and surface effects even in the limit of a soft symmetry energy, while $b$ quantifies the sensitivity of the neutron skin to changes in the slope parameter $L$.

Microscopic energy density functional (EDF) calculations provide strong support for this linear correlation. Systematic studies using large sets of Skyrme and relativistic mean-field interactions have shown that, for heavy nuclei such as $^{208}$Pb, the neutron skin thickness exhibits a nearly linear dependence on $L$ over a wide range of models. Typical values found in the literature are \cite{RocaMaza2011,Reinhard2010,Horowitz2014,RocaMaza2015,Chen2005,Vinas2014}
$
\Delta r_{np}(^{208}\mathrm{Pb}) \approx (0.10-0.15)\,\text{fm}
+ (0.001-0.002)\,L,
$
with $L$ expressed in MeV.
This behavior can be traced back to the fact that both $\Delta r_{np}$ and $L$ are governed by the same underlying quantity, namely the pressure difference between neutrons and protons in asymmetric matter at densities around $\rho_0$. Since the neutron skin probes the surface region of nuclei, where densities are slightly below saturation, it is particularly sensitive to the slope of the symmetry energy rather than its absolute value at $\rho_0$.
An important consequence of this correlation is that measurements of neutron skin thickness in finite nuclei provide direct constraints on the EOS of neutron-rich matter. Conversely, astrophysical observations that constrain $L$, such as neutron star radii or tidal deformabilities, can be used to infer neutron skin properties.
Therefore, the neutron skin thickness serves as a powerful bridge connecting nuclear structure, reaction dynamics, and astrophysical phenomena, enabling a unified understanding of isovector properties across vastly different physical systems.

\section{Dipole Polarizability}

The electric dipole polarizability $\alpha_D$ is a fundamental observable characterizing the response of a nucleus to an external electric field. It is defined in terms of the photoabsorption cross section $\sigma_\gamma(E)$ as
\begin{equation}
\alpha_D = \frac{\hbar c}{2\pi^2} \int_0^\infty \frac{\sigma_\gamma(E)}{E^2} \, dE.
\end{equation}
Physically, $\alpha_D$ measures how easily the neutron and proton distributions can be displaced relative to each other. This collective oscillation corresponds to the isovector giant dipole resonance (GDR), where neutrons oscillate against protons. Because of the inverse energy weighting in the integral, $\alpha_D$ is particularly sensitive to low-energy dipole strength, including contributions from the so-called pygmy dipole resonance (PDR), which is often associated with oscillations of the neutron skin against the nuclear core.

A deep connection between $\alpha_D$ and the neutron skin thickness $\Delta r_{np}$ emerges from both microscopic energy density functional (EDF) calculations and macroscopic models such as the droplet model. In neutron-rich nuclei, the symmetry energy governs the restoring force of the dipole oscillation. A softer symmetry energy (smaller $L$) leads to a stronger restoring force and therefore a smaller polarizability, whereas a stiffer symmetry energy (larger $L$) results in a weaker restoring force and a larger $\alpha_D$.
Within the droplet model, one finds an approximate expression for the dipole polarizability \cite{RocaMaza2013,Piekarewicz2012}
\begin{equation}
\alpha_D \approx \frac{\pi e^2}{54} \frac{A \langle r^2 \rangle}{J}
\left( 1 + \frac{5}{3}\frac{L}{J}\epsilon_A \right),
\end{equation}
where $J$ is the symmetry energy at saturation density, $\langle r^2 \rangle$ is the mean-square radius, and $\epsilon_A$ accounts for deviations of the nuclear density from saturation. This expression shows explicitly that $\alpha_D$ is inversely proportional to $J$ and increases with the slope parameter $L$. At the same time, the neutron skin thickness can also be expressed in terms of symmetry energy parameters, as discussed in the previous section. Since both $\alpha_D$ and $\Delta r_{np}$ depend on the same underlying isovector properties of the EOS, a strong correlation between these observables naturally arises.
Indeed, extensive EDF calculations have demonstrated that $\alpha_D$ and $\Delta r_{np}$ are nearly linearly correlated for heavy nuclei. This correlation can be expressed schematically as
$
\Delta r_{np} \simeq a' + b'\,\alpha_D,
$
or, more robustly, through the product $\alpha_D J$, which exhibits an even tighter correlation with $\Delta r_{np}$,
$
\alpha_D J \propto \Delta r_{np}.
$

The improved correlation with $\alpha_D J$ arises because the dependence on $J$ is largely factored out, reducing model-dependent uncertainties. This insight has been confirmed by covariance analyses of EDF parameter sets, which show that the correlation coefficient between $\alpha_D J$ and $\Delta r_{np}$ is close to unity for heavy nuclei such as $^{208}$Pb.
From a physical perspective, the connection between $\alpha_D$ and $\Delta r_{np}$ can be understood as follows. A larger neutron skin implies a more extended neutron distribution, which reduces the restoring force for isovector oscillations and enhances the dipole response. Consequently, nuclei with thicker neutron skins exhibit larger dipole polarizabilities.
Experimentally, high-resolution measurements of $\alpha_D$, particularly via proton inelastic scattering at forward angles or polarized photon scattering, have provided precise constraints on neutron skin thickness \cite{Tamii2011,Endres2010,Adrich2005,Klimkiewicz2007,Wieland2009,Rossi2013,Bracco2019} For example, measurements of $\alpha_D$ in $^{208}$Pb.
Therefore, the dipole polarizability serves as a powerful and complementary observable for constraining the density dependence of the symmetry energy. When combined with other measurements, such as neutron skins and astrophysical observations, it provides a coherent and consistent picture of the EOS of neutron-rich matter.

\section{Parity-Violating Electron Scattering}

The parity-violating (PV) electron scattering  method exploits the interference between electromagnetic and weak neutral currents in electron--nucleus scattering, leading to a small but measurable asymmetry in the cross section for right- and left-handed electrons. It is defined as
$
A_{\rm PV} = ({d\sigma_R - d\sigma_L})/({d\sigma_R + d\sigma_L}),
$
where $d\sigma_{R(L)}$ denotes the differential cross section for right- (left-) handed electrons. In the Born approximation, this asymmetry can be written as
\begin{equation}
A_{\rm PV} \simeq \frac{G_F Q^2}{4\pi \alpha \sqrt{2}} 
\frac{F_W(Q^2)}{F_{\rm ch}(Q^2)},
\end{equation}
where $G_F$ is the Fermi constant, $\alpha$ is the fine-structure constant, $Q^2$ is the squared momentum transfer, $F_{\rm ch}(Q^2)$ is the electromagnetic (charge) form factor, and $F_W(Q^2)$ is the weak form factor.
The crucial feature of this observable is that the weak charge of the neutron is much larger than that of the proton. As a result, the weak form factor is dominated by the neutron distribution,
\begin{equation}
F_W(Q^2) \approx \int d^3r \, \frac{\sin(Qr)}{Qr} \, \rho_n(r),
\end{equation}
to leading order. In contrast, the electromagnetic form factor $F_{\rm ch}(Q^2)$ is primarily sensitive to the proton distribution $\rho_p(r)$.
At low momentum transfer, the form factors can be expanded as
$
F(Q^2) \approx 1 - {Q^2}/{6} \langle r^2 \rangle + \cdots,
$
so that the PV asymmetry becomes sensitive to the difference between neutron and proton mean-square radii. To leading order, one obtains
$
A_{\rm PV} \propto Q^2 \left( 1 - ({Q^2}\langle r_n^2 - r_p^2 \rangle/6 \right),
$
which shows explicitly that the asymmetry is directly related to the neutron skin thickness.
A more transparent relation can be obtained by expressing the weak radius $R_W$ in terms of neutron and proton distributions. One finds that the difference between the weak and charge radii is approximately proportional to the neutron skin thickness,
$
R_W - R_{\rm ch} \propto \Delta r_{np}.
$

Therefore, measurements of $A_{\rm PV}$ at fixed momentum transfer allow a direct extraction of $\Delta r_{np}$ with minimal theoretical uncertainties, limited mainly by electroweak corrections and small contributions from strange quark form factors.
From a physical standpoint, the sensitivity of PV scattering to the neutron skin arises because the weak interaction effectively ``filters out'' the neutron distribution. This makes PV electron scattering uniquely suited to probe the spatial extent of neutrons in nuclei, in contrast to hadronic probes that suffer from strong interaction uncertainties \cite{PREX,CREX,PREX2,Horowitz2012,RocaMaza2011}.
Experimentally, the PREX and CREX experiments at Jefferson Lab have provided landmark measurements of the parity-violating asymmetry in $^{208}$Pb and $^{48}$Ca, respectively. These measurements have yielded direct determinations of neutron skin thicknesses. For example, PREX-II reports
$
\Delta r_{np}(^{208}\mathrm{Pb}) = 0.283 \pm 0.071 \, \mathrm{fm},
$
while CREX finds
$
\Delta r_{np}(^{48}\mathrm{Ca}) = 0.121 \pm 0.026 \, \mathrm{fm}.
$


\section{ Nuclear Fragmentation Observables}

At intermediate and high energies, nuclear reactions can be described within the Glauber multiple-scattering framework, which provides a powerful tool to connect reaction observables with nuclear density distributions. In this approach, the elastic S-matrix for a collision at impact parameter ${\bf b}$ is written as
$
S({\bf b}) = \exp\left[i\chi(\bf b)\right],
$
where the eikonal phase $\chi(\bf{b})$ encodes the cumulative effect of nucleon-nucleon interactions.
For nucleus-nucleus collisions, the phase can be expressed in terms of projectile and target densities,
\begin{equation}
\chi({\bf b}) = - \int d^2s \, \Gamma_{NN}({\bf b}-{\bf s}) 
\int dz \, \rho_P({\bf s},z) \int dz' \, \rho_T({\bf s},z'),
\end{equation}
where $\Gamma_{NN}$ is the profile function related to the nucleon-nucleon scattering amplitude, and $\rho_{P,T}$ are the projectile and target densities.
The total reaction cross section is given by
$
\sigma_R = \int d^2b \, \left[1 - |S(\bf{b})|^2\right].
$
While $\sigma_R$ depends on the overall size of the nuclear density distribution, it is only weakly sensitive to the detailed structure of the neutron skin. This can be understood by noting that $\sigma_R$ is dominated by the geometrical overlap of the nuclear densities, which scales approximately with the square of the interaction radius. As a consequence, moderate changes in the neutron skin thickness lead to relatively small variations in $\sigma_R$.

In contrast, fragmentation observables, and in particular neutron-removal cross sections, exhibit a much stronger sensitivity to the neutron density distribution in the nuclear surface region. Within the Glauber model, the cross section for removing one or more neutrons from the projectile can be written schematically as
\begin{equation}
\sigma_{\Delta N} = \int d^2b \, P_{\rm surv}^{(p)}(b)\left[1 - P_{\rm surv}^{(n)}(b)\right],
\end{equation}
where $P_{\rm surv}^{(p)}$ and $P_{\rm surv}^{(n)}$ are the survival probabilities of protons and neutrons, respectively.
The neutron survival probability depends on the neutron thickness function,
\begin{equation}
T_n(b) = \int dz \, \rho_n\left(\sqrt{b^2+z^2}\right),
\end{equation}
which emphasizes the importance of the neutron density at large radii. Since the neutron skin corresponds precisely to an excess of neutrons in the surface region, neutron-removal cross sections become highly sensitive to $\Delta r_{np}$.

This sensitivity can be understood more quantitatively by considering small variations of the neutron density profile. A change in the neutron skin thickness $\Delta r_{np}$ modifies the tail of the neutron density, leading to a change in the thickness function $T_n(b)$ at large impact parameters. Because fragmentation processes are dominated by peripheral collisions, the corresponding cross sections scale approximately with the surface neutron density,
\begin{equation}
\delta \sigma_{\Delta N} \propto \int d^2b \, \delta T_n(b),
\end{equation}
which leads to an enhanced response to $\Delta r_{np}$.
Extensive calculations using realistic density distributions have demonstrated that, along isotopic chains such as Sn isotopes, the neutron skin thickness can vary by $\sim 0.2$ fm, while the total reaction cross section changes by only a few percent. In contrast, neutron-removal cross sections can change by $\sim 20\%$, providing a significantly more sensitive observable.

This behavior has been confirmed in systematic studies of fragmentation reactions using Glauber and eikonal models \cite{Aumann2017,Teixeira2022,Bertulani2020}. These works show that neutron-removal cross sections exhibit a nearly monotonic dependence on $\Delta r_{np}$, allowing for the extraction of neutron skin thicknesses from experimental data.
More recently, the formalism has been extended to include dynamical effects and correlations, as well as applications to high-energy collisions at RHIC and the LHC \cite{Bertulani2019,Bertulani2025}. These studies emphasize that fragmentation reactions, when properly analyzed within the Glauber framework, provide a robust and complementary probe of neutron skins.


\section{Ultraperipheral Collisions and Neutron Skin Sensitivity}

Ultraperipheral collisions (UPCs) of relativistic heavy ions provide a powerful electromagnetic probe of nuclear structure at large impact parameters, where the nuclei interact predominantly through their electromagnetic fields rather than through strong interactions. In this regime, the interaction can be described using the equivalent photon method (EPM).
Within this framework, the cross section for a given electromagnetic process can be written as \cite{BB88}
\begin{equation}
\sigma = \int \frac{d\omega}{\omega} \, n(\omega) \, \sigma_\gamma(\omega),
\end{equation}
where $n(\omega)$ is the equivalent photon number (EPN) and $\sigma_\gamma(\omega)$ is the photonuclear cross section at photon energy $\omega$. The EPN depends on the charge distribution of the projectile nucleus and is strongly peaked at low photon energies.
For a relativistic nucleus with Lorentz factor $\gamma$, the photon spectrum can be expressed as
\begin{equation}
n(\omega,b) \propto Z^2 \left| \int d^2q_\perp \frac{F(q)}{q_\perp^2 + \omega^2/\gamma^2} e^{i {\bf q}_\perp \cdot {\bf b}} \right|^2,
\end{equation}
where $F(q)$ is the nuclear form factor, given by the Fourier transform of the charge density.

Although the photon flux is primarily determined by the proton distribution, the photonuclear response $\sigma_\gamma(\omega)$ is sensitive to the neutron distribution through isovector excitations. In particular, the excitation of the giant dipole resonance (GDR) and the pygmy dipole resonance (PDR) involves oscillations of neutrons against protons, making these processes sensitive to the neutron skin thickness.
The photonuclear cross section can be decomposed into multipole contributions,
\begin{equation}
\sigma_\gamma(\omega) = \sum_{E/M,L} \sigma_\gamma^{(E/M,L)}(\omega),
\end{equation}
with the dominant contribution in heavy nuclei arising from the electric dipole ($E1$) response. As discussed previously, the $E1$ strength distribution is strongly correlated with the neutron skin thickness.

A direct connection between UPC observables and the neutron skin can be established by noting that the $E1$ response function depends on the spatial separation between neutron and proton distributions. In particular, the dipole strength function can be written schematically as
$
S_{E1}(\omega) \propto \left| \langle \Psi_f | \hat{D} | \Psi_i \rangle \right|^2,
$
where the dipole operator is
$
\hat{D} = ({N}/{A}) \sum_{p} {\bf r}_p - ({Z}/{A}) \sum_{n} {\bf r}_n.
$
This operator explicitly involves the relative displacement of proton and neutron centers of mass, and therefore depends directly on the neutron skin thickness.
Consequently, the integrated photonuclear cross section, and hence the UPC cross section, exhibits a dependence on $\Delta r_{np}$. To leading order, one can write
\begin{equation}
\sigma_{\rm UPC}^{(E1)} \propto \int \frac{d\omega}{\omega} n(\omega) S_{E1}(\omega)
\sim \alpha_D,
\end{equation}
which connects UPC observables directly to the dipole polarizability. Since $\alpha_D$ is correlated with the neutron skin thickness, as discussed earlier, one obtains an approximate relation
$
\sigma_{\rm UPC} \simeq a + b \, \Delta r_{np},
$
where the coefficients $a$ and $b$ depend on the collision system and beam energy.

In addition to electromagnetic excitation, UPCs also allow access to coherent vector meson production, such as $J/\psi$ photoproduction, also at the Electron Ion Collider (EIC) \cite{Bak25}. In these processes, the cross section depends on the nuclear gluon distribution and is sensitive to the nuclear geometry through the thickness function
$
T(b) = \int dz \, \rho(\sqrt{b^2+z^2}),
$
which encodes the spatial distribution of matter. Variations in the neutron skin modify the overall density profile and can therefore affect coherent production cross sections.

Recent theoretical studies have demonstrated that UPC observables, including neutron emission probabilities and vector meson (e.g., J/$\Psi$) production cross sections, exhibit measurable sensitivity to neutron skin thickness \cite{Bertulani2020,Bak25}. In particular, the probability of electromagnetic dissociation via GDR excitation is enhanced for nuclei with larger neutron skins, due to the increased dipole strength.
From a physical perspective, UPCs probe the nuclear surface region, where the electromagnetic fields are strongest and where the neutron skin resides. This makes UPCs complementary to hadronic probes such as fragmentation reactions, and to electroweak probes such as parity-violating electron scattering.
Therefore, UPCs provide a unique and versatile tool for studying neutron skins, connecting nuclear structure, reaction dynamics, and high-energy heavy-ion physics within a unified framework.

\section{Quasi-Free Scattering and Sensitivity to the Neutron Skin}

Quasi-free scattering reactions, such as $(p,2p)$ and $(p,pn)$, provide a powerful tool to probe single-particle structure and the spatial distribution of nucleons in nuclei. At sufficiently high energies, these reactions can be described within the impulse approximation, where the incoming proton interacts predominantly with a single nucleon inside the target nucleus, while the remaining nucleons act as spectators.
Within this framework, the triple-differential cross section can be written schematically as \cite{Jacob1966}
\begin{equation}
\frac{d^3\sigma}{dE_1 d\Omega_1 d\Omega_2}
\propto K \, S(p_m,E_m) \, \frac{d\sigma_{NN}}{d\Omega},
\end{equation}
where $K$ is a kinematic factor, $S(p_m,E_m)$ is the spectral function describing the momentum and energy distribution of the struck nucleon, and $d\sigma_{NN}/d\Omega$ is the elementary nucleon-nucleon cross section. The missing momentum $p_m$ is defined as the difference between the initial and final momenta, and encodes information about the nuclear wave function.
At intermediate energies, distortions due to initial- and final-state interactions (ISI/FSI) must be included. This is typically achieved within the distorted-wave impulse approximation (DWIA) or its eikonal counterpart, where the scattering amplitude is modified by attenuation factors depending on the nuclear density.
The sensitivity of quasi-free scattering to neutron skins arises primarily through two mechanisms: (i) the spatial dependence of the nucleon knockout probability, and (ii) spin-dependent interference effects in the scattering amplitude.
The knockout probability is strongly influenced by absorption effects, which suppress contributions from the nuclear interior. As a result, quasi-free scattering is predominantly sensitive to nucleons located near the nuclear surface. This surface dominance implies that variations in the neutron density at large radii, i.e., the neutron skin, can significantly affect the measured observables.

A more subtle and particularly sensitive observable is the analyzing power $A_y$, which arises from the spin dependence of the nucleon--nucleon interaction. In $(p,2p)$ reactions, the analyzing power can be expressed as
$
A_y = ({d\sigma(\uparrow) - d\sigma(\downarrow)})/
({d\sigma(\uparrow) + d\sigma(\downarrow)}),
$
where $\uparrow$ and $\downarrow$ denote the spin orientation of the incoming proton.
The connection to the neutron skin emerges through the so-called Maris polarization effect \cite{Maris1979}. This effect originates from the interference between scattering amplitudes corresponding to different spin-orbit partners (e.g., $p_{1/2}$ and $p_{3/2}$ states), combined with the asymmetry between near-side and far-side scattering paths. The effective analyzing power difference between these orbitals can be written as
$
\Delta A_y = A_y(p_{1/2}) - A_y(p_{3/2}),
$
which is particularly sensitive to the spatial distribution of nucleons.
In neutron-rich nuclei, the presence of a neutron skin modifies the absorption and scattering probabilities differently for near-side and far-side trajectories. Since neutrons dominate the surface region, the spin-dependent part of the nucleon--nucleon interaction is affected, leading to measurable changes in $\Delta A_y$.
Microscopic calculations have shown that $\Delta A_y$ exhibits a strong and approximately linear dependence on the neutron skin thickness,
$
\Delta A_y \simeq a + b \, \Delta r_{np},
$
where the coefficients $a$ and $b$ depend on the reaction kinematics and the specific orbitals involved. This relation arises because both $\Delta A_y$ and $\Delta r_{np}$ are controlled by the neutron density in the surface region, which determines the strength of absorption and the phase differences between competing amplitudes.

In addition to analyzing powers, integrated knockout cross sections also show sensitivity to the neutron skin. The attenuation factor in the eikonal approximation can be written as
$
S(b) = \exp\left[-\sigma_{NN} \int dz \, \rho(\sqrt{b^2+z^2}) \right],
$
which depends explicitly on the nuclear density profile. Variations in the neutron skin alter the density at large radii, thereby modifying the survival probability and the overall cross section.
Systematic studies of quasi-free scattering on isotopic chains, particularly Sn isotopes, have demonstrated that both $\Delta A_y$ and knockout cross sections provide complementary constraints on neutron skin thickness \cite{Shubchintak2018,Bertulani2026}. These observables are especially valuable because they probe different aspects of the nuclear density: while cross sections are sensitive to overall absorption, analyzing powers encode detailed spin-dependent dynamics.

\section{Bayesian Analysis and Constraints on the Symmetry Energy}

A quantitative extraction of neutron skin thicknesses and symmetry energy parameters from experimental data requires a statistical framework capable of combining heterogeneous datasets and theoretical models. Bayesian inference provides such a framework, allowing for the consistent propagation of uncertainties and the incorporation of prior knowledge.
One seeks to determine the posterior probability distribution of a set of parameters ${\bf \theta}$, which may include symmetry energy coefficients such as $J$ and $L$, as well as nuisance parameters describing systematic uncertainties. The posterior distribution is given by Bayes' theorem,
$
P({\bf \theta}|\mathcal{D}) = {P(\mathcal{D}|{\bf \theta}) P({\bf \theta})}/{P(\mathcal{D})},
$
where $\mathcal{D}$ denotes the experimental data, $P(\mathcal{D}|{\bf \theta})$ is the likelihood function, $P({\bf \theta})$ is the prior distribution, and $P(\mathcal{D})$ is the evidence.

In the context of neutron skin studies, the data $\mathcal{D}$ consist of measurements from multiple probes, including dipole polarizabilities, parity-violating asymmetries, fragmentation cross sections, and quasi-free scattering observables. Each dataset provides constraints on $\Delta r_{np}$, which in turn depends on the underlying EOS parameters.
Following the methodology developed in recent work by Azizi, Bertulani, and Davila \cite{Azizi2026}, the neutron skin thickness across a range of nuclei can be parameterized as
$
\Delta r_{np} = \beta_0 + \beta_1 I + \beta_2 A^{-1/3} + \beta_3 I A^{-1/3},
$
where $I = (N-Z)/A$ is the isospin asymmetry, and the coefficients $\beta_i$ encode the dependence on nuclear structure and the symmetry energy.
The likelihood function is constructed by comparing model predictions $\Delta r_{np}^{\rm th}$ with experimental values $\Delta r_{np}^{\rm exp}$,
\begin{equation}
P(\mathcal{D}|{\bf \theta}) \propto \exp\left[-\frac{1}{2}
\sum_i \frac{\left(\Delta r_{np,i}^{\rm exp} - \Delta r_{np,i}^{\rm th}({\bf \theta})\right)^2}
{\sigma_i^2 + \tau^2}
\right],
\end{equation}
where $\sigma_i$ are experimental uncertainties and $\tau$ represents an additional model discrepancy parameter accounting for systematic effects.

A key feature of the analysis is the incorporation of predictions from a large set of energy density functionals (EDFs), including both Skyrme and relativistic mean-field models. These EDFs provide theoretical mappings between neutron skin thickness and symmetry energy parameters, effectively serving as an emulator for the nuclear equation of state. The posterior distribution can therefore be projected onto the $(J,L)$ plane, yielding constraints on the symmetry energy and its slope.
In practice, the inference is performed using Markov Chain Monte Carlo (MCMC) techniques, which sample the posterior distribution and allow for the extraction of credible intervals. The resulting constraints $
J = J_0 \pm \Delta J,$ and 
$L = L_0 \pm\Delta J,
$ where the uncertainties reflect both experimental errors and model uncertainties.

An important outcome of this Bayesian framework is the ability to combine diverse experimental probes in a unified analysis. For example, neutron skin values extracted from dipole polarizability measurements can be combined with those obtained from parity-violating electron scattering and fragmentation reactions, leading to a more precise determination of EOS parameters than any single probe alone.
Furthermore, the Bayesian approach allows for the identification of correlations between observables. In particular, strong correlations are found between $\Delta r_{np}$, the dipole polarizability $\alpha_D$, and the slope parameter $L$. These correlations can be quantified through covariance matrices or posterior correlation coefficients, providing insight into the underlying physics.
Another advantage of the method is its ability to propagate uncertainties to astrophysical observables. Once the posterior distribution for $(J,L)$ is obtained, it can be used as input for neutron star structure calculations via the TOV equations. This enables predictions for neutron star radii and tidal deformabilities consistent with nuclear physics constraints.
The analysis of Azizi, Bertulani, and Davila \cite{Azizi2026} demonstrates that combining neutron skin data with EDF predictions leads to well-constrained values of the symmetry energy parameters, with typical results around $J \sim 32$ MeV and $L \sim 45-50$ MeV. These values are consistent with independent constraints from nuclear experiments and astrophysical observations, reinforcing the robustness of the Bayesian approach.

\section*{Acknowledgments}

This work was supported by U.S. Department of Energy Office of Nuclear Physics under Contract No. DE-SC0026074 with East Texas A\&M University.

\end{document}